\documentclass{article}
\usepackage{dcase2021_techrep,amsmath,graphicx,url,times,booktabs, tabularx, multirow, color, comment}

\title{Ensemble of ACCDOA- and EINV2-based Systems with D3Nets and Impulse Response Simulation for Sound Event Localization and Detection}

\name{Kazuki Shimada, Naoya Takahashi, Yuichiro Koyama, Shusuke Takahashi,}
\secondlinename{Emiru Tsunoo, Masafumi Takahashi, Yuki Mitsufuji}
\address{Sony Group Corporation, Japan}

\begin{document}

\ninept
\maketitle

\begin{sloppy}

\begin{abstract}
This report describes our systems submitted to the DCASE2021 challenge task~3: sound event localization and detection~(SELD) with directional interference. Our previous system based on activity-coupled Cartesian direction of arrival~(ACCDOA) representation enables us to solve a SELD task with a single target. This ACCDOA-based system with efficient network architecture called RD3Net and data augmentation techniques outperformed state-of-the-art SELD systems in terms of localization and location-dependent detection. Using the ACCDOA-based system as a base, we perform model ensembles by averaging outputs of several systems trained with different conditions such as input features, training folds, and model architectures. We also use the event independent network v2~(EINV2)-based system to increase the diversity of the model ensembles. To generalize the models, we further propose impulse response simulation~(IRS), which generates simulated multi-channel signals by convolving simulated room impulse responses~(RIRs) with source signals extracted from the original dataset. Our systems significantly improved over the baseline system on the development dataset.

\end{abstract}


\section{Introduction}
\label{sec:intro}
Sound event localization and detection~(SELD) involves identifying both the direction of arrival~(DOA) and the type of sound. Numerous methods have been tackling SELD through the DCASE challenge~\cite{politis2021dataset,politis2020dataset,wang2021four,cao2021improved,nguyen2021general,shimada2021accdoa}.
The SELD system based on activity-coupled Cartesian direction of arrival~(ACCDOA) representation enables us to solve a SELD task with a single target~\cite{shimada2021accdoa}. This ACCDOA-based system with RD3Net~\cite{Takahashi20d3netCVPR,shimada2021accdoa} and data augmentation techniques outperformed state-of-the-art SELD systems in terms of localization and location-dependent detection~\cite{shimada2021accdoa}.

% In this study, we used the ACCDOA-based system as a base and performed model ensembles by averaging outputs of several systems trained with different conditions such as input features, training folds, and model architectures. To further increase the diversity of the model ensemble, we also used the event independent network v2~(EINV2)-based system~\cite{cao2021improved}.
% To increase the training data, we use equalized mixture data augmentation~(EMDA), rotation of first-order Ambisonic~(FOA) signals, and the multichannel version of SpecAugment. In addition to these augmentation techniques, we further propose impulse response simulation~(IRS), which generates simulated multi-channel signals by convolving simulated room impulse responses~(RIRs) with source signals extracted from the original dataset.
% Experiments on the development dataset showed our system significantly improve over the baseline system.
In this study, we perform model ensembles of several systems trained with different conditions and model architectures.
We use ACCDOA-based systems as a base and perform model ensembles by averaging outputs of the systems trained with various input features, training folds, and model architectures.
The event independent network v2~(EINV2)-based system~\cite{cao2021improved} is also used to increase the diversity of the model architectures.
To increase the training data, we carry out four kinds of data augmentation: 1) equalized mixture data augmentation~\cite{Takahashi16,Takahashi2017AENet}, 2) rotation of first-order Ambisonic~(FOA) signals~\cite{mazzon2019first}, 3) the multichannel version of SpecAugment~\cite{park2019specaugment,shimada2021accdoa}, and 4) impulse response simulation~(IRS), which is a novel approach for SELD to generates simulated multi-channel signals by convolving simulated room impulse responses~(RIRs) with source signals extracted from the original dataset.
Experiments on the development dataset showed our system significantly improved over the baseline system.

\section{ACCDOA-based System}
\label{sec:accdoa_system}
In this section, we first give the framework of ACCDOA-based systems. Then we explain the parts of our pipeline: the features, data augmentation, and network architecture.

\begin{figure}[t]
    \centering
    \centerline{\includegraphics[width=0.92\linewidth]{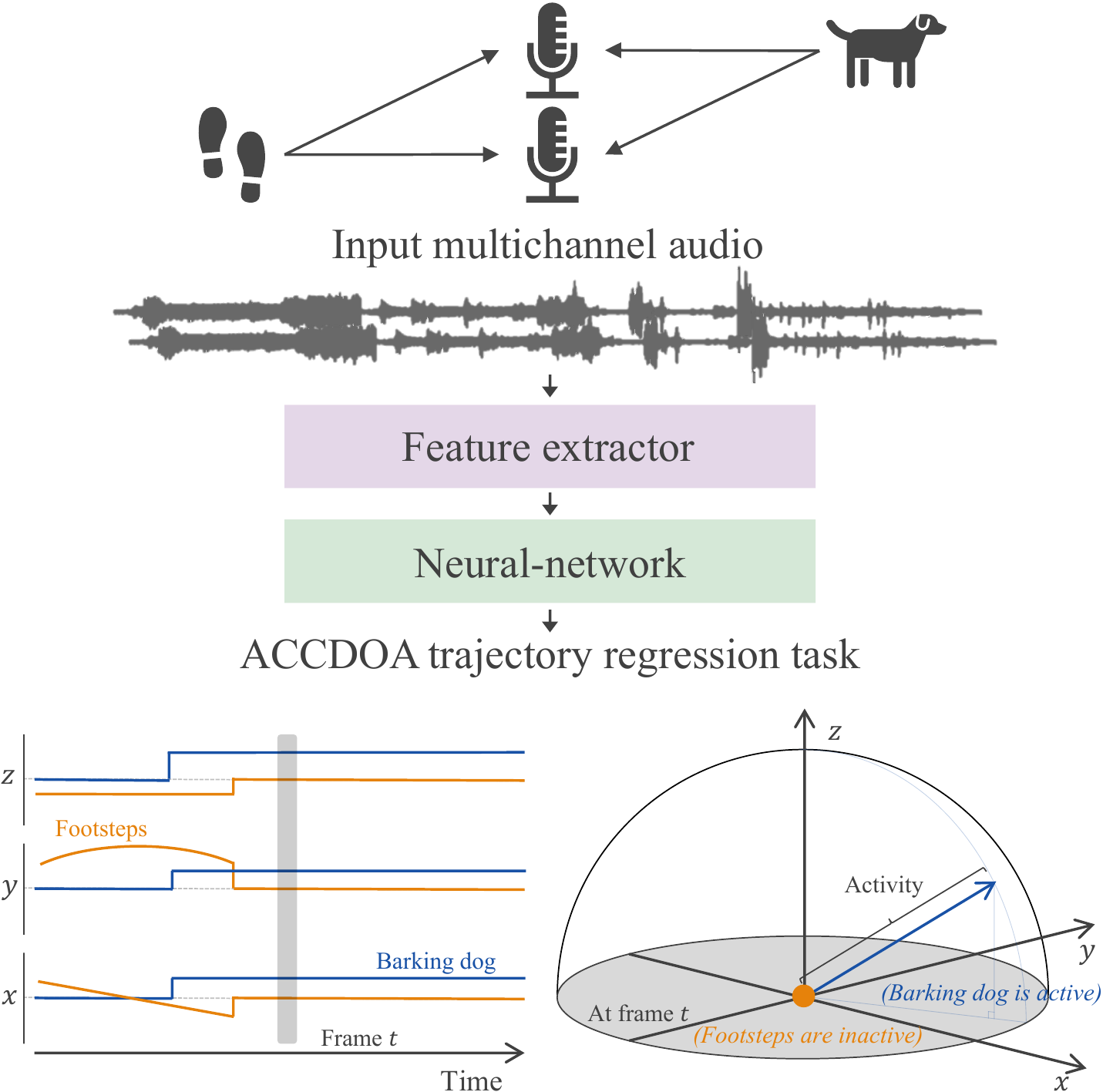}}
    \caption{Illustration of an ACCDOA-based SELD system.}
    \label{fig:overview}
\end{figure}

\subsection{ACCDOA framework}
\label{ssec:accdoa_framework}
The ACCDOA representation assigns a sound event activity to the length of the corresponding Cartesian DOA vector, which enables us to handle SELD as a single task with a single network~\cite{shimada2021accdoa}. A schematic flow of the ACCDOA-based SELD system is shown in Fig.~\ref{fig:overview}. 
After extracting the features, the network outputs frame-wise ACCDOA vectors for target sound events. The model is trained to minimize the Euclidean distance between the estimated and target coordinates in the ACCDOA representation. We solve the multi-output regression with a mean square error~(MSE) loss~\cite{shimada2021accdoa}.

\subsection{Features}
\label{ssec:features}
Multichannel amplitude spectrograms and inter-channel phase differences~(IPDs) are used as frame-wise features. Here, ${\rm IPD}_{t, f, p, q} = \angle{x}_{t, f, p} - \angle{x}_{t, f, q}$ is computed from the short-time Fourier transform~(STFT) coefficients $x_{t, f ,p}$ and $x_{t, f, q}$, where $t, f, p,$ and $q$ denote the time frame, the frequency bin, the Ambisonic channel $p$, and the channel $q$, respectively. We fix $p=0$ to compute relative IPDs between all the other channels, $q\neq0$. Since the input consists of four Ambisonic channel signals, we can extract four amplitude spectrograms and three IPDs.
Some of the models use PCEN~\cite{lostanlen2018per}, cosIPDs, and sinIPDs~\cite{wang2018multi} as input features, instead of the amplitude spectrograms and IPDs.

\subsection{Data augmentation}
\label{ssec:augmentation}
To increase the generalizability of the model, we use three data augmentation techniques as described in our previous paper~\cite{shimada2021accdoa}: EMDA~\cite{Takahashi16,Takahashi2017AENet}, rotation of FOA signals~\cite{mazzon2019first}, and the multichannel version of SpecAugment~\cite{park2019specaugment,shimada2021accdoa}.
We use the EMDA method, where up to two audio events are mixed with random amplitudes, delays, and the modulation of frequency characteristics, i.e., equalization~\cite{Takahashi16,Takahashi2017AENet}.
We also use the spatial augmentation method~\cite{mazzon2019first}. It rotates the training data represented in the FOA format and enables us to increase the numbers of DOA labels without losing the physical relationships between steering vectors and observations.
Lastly, we use the multichannel version of SpecAugment~\cite{park2019specaugment,shimada2021accdoa}. SpecAugment was extended to the channel dimension in addition to the time-frequency hard masking schemes applied on amplitude spectrograms~\cite{shimada2021accdoa}.

In addition to these augmentation techniques, we further propose IRS.
% , which generates simulated multi-channel signals by convolving simulated RIRs with source signals extracted from the original dataset.
Fig.~\ref{fig:irs_augmentation} describes the workflow of IRS augmentation. While source signals are extracted from the original reverberant dataset, RIRs for the FOA format are simulated. Then the simulated RIRs are convloved with the extracted source signals to obtain augmented multi-channel signals for the training dataset.

To extract source signals from the original dataset, we first extract non-overlapping and non-moving sound event segments from the original dataset on the basis of the provided sound event detection~(SED) label. Since the dataset of DCASE2021 task3 contains unlabeled directional interference events, we heuristically eliminate the interference events on the basis of a two-stage approach. In the first stage, the sound event segments are eliminated when the segments are not recognized as any sound event classes by a preliminarily trained model. These segments can be regarded as the ones that dominantly include interference events. In the second stage, using the covariance matrix of the segment in the frequency bins under 2 kHz and their eigenvalues, the segments are removed when the second eigenvalue is larger than 30\% of the first eigenvalue in most frequency bins. Finally, complex Gaussian mixture model~(CGMM)-based minimum variance distortionless response~(MVDR) beamforming is applied to the extracted segments to obtain the source signals of the target sound events~\cite{higuchi2016, wang2021four}.

In order to obtain RIRs for the FOA format, 
RIRs for the Eigenmike are first simulated assuming that the Eigenmike can be regarded as a rigid baffle array~\cite{Politis2016, MitsufujiTKS21}. An image source method~\cite{Scheibler2018} is utilized for simulating reverberant conditions. The reverberation time~(RT60) is randomly defined to be within 100 to 500~ms. Then the simulated RIRs are converted to FOA format on the basis of a higher-order Ambisonics~(HOA) encoding process~\cite{moreau2006, Archontis2017}. 

% Finally, the simulated RIRs are convolved with the source signals, thus we obtain augmented multi-channel signals that are used for the training dataset. 

\begin{figure}[t]
    \centering
    \centerline{\includegraphics[width=0.66\linewidth]{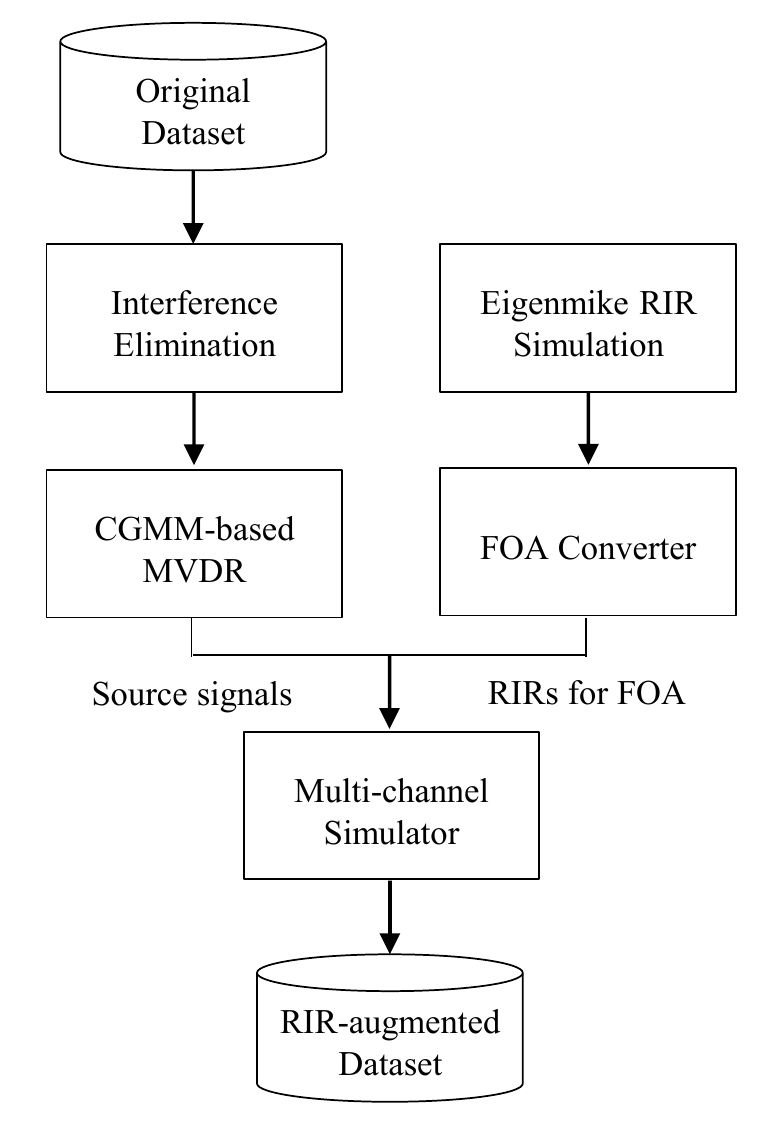}}
    \caption{The workflow of IRS.}
    \label{fig:irs_augmentation}
    \vspace{-2mm}
\end{figure}

\subsection{Network architecture}
\label{ssec:network}

\begin{figure*}[t]
    \centering
    \centerline{\includegraphics[width=0.99\linewidth]{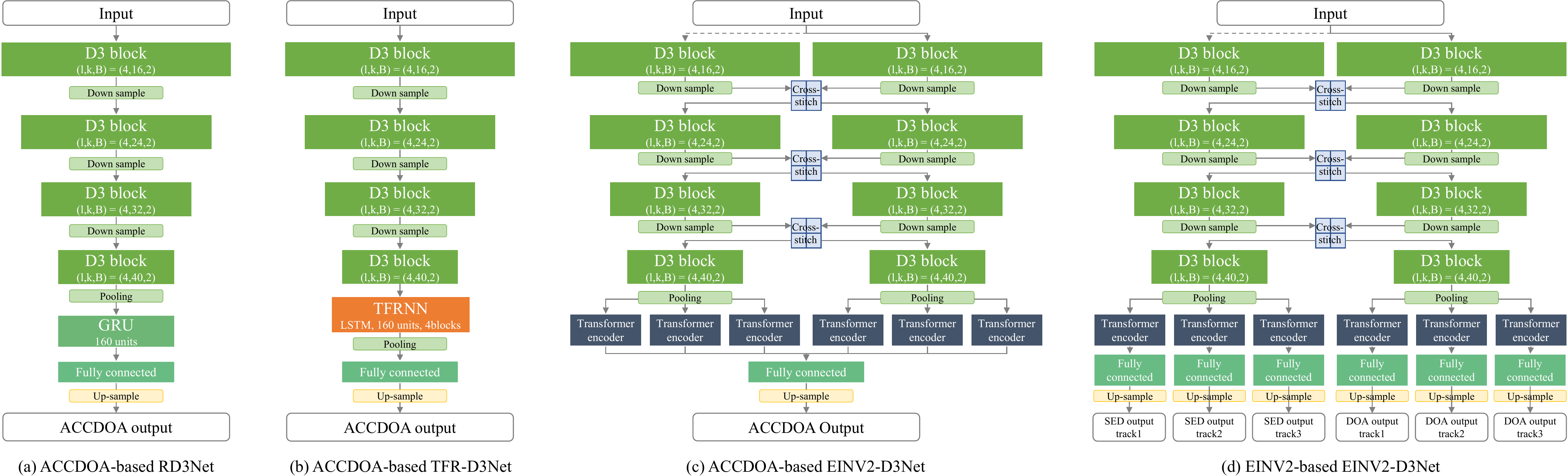}}
    \caption{Illustration of D3Net architectures. (a)~ACCDOA-based RD3Net, (b)~ACCDOA-based TFR-D3Net (c)~ACCDOA-based EINV2-D3Net (d)~EINV2-based EINV2-D3Net. $l, k$, and $B$ in D3 blocks denote the number of layers, growth rate, and number of blocks, respectively. }
    \label{fig:netarch}
\end{figure*}

% \begin{comment}
% As the network architecture, we adopt the D3Net architecture~\cite{Takahashi20d3netCVPR}, which has achieved the state-of-the-art performance in music source separation. To adapt D3Net architecture to the SELD problem, three modifications are made. First, dense blocks in the up-sampling path is omitted because high frame-rate prediction is not necessary for the SELD problem. Second, the bottleneck part is replaced with gated recurrent unit~(GRU) cells. Third, the batch normalization is replaced with the network deconvolution~\cite{Ye2020}. 
% % In each dense block, the dilation factor of the initial convolution is set to one, and it doubles every time the next convolution is applied, as applied in WaveNet~\cite{Aaron2016WN}. 
% This architecture is called \emph{RD3Net}~\cite{shimada2021accdoa}; and is illustrated in Fig.~\ref{fig:netarch}~(a).
% \end{comment}

In this study, to increase the diversity of the model ensemble, we consider three variants of D3Net architecture~\cite{Takahashi20d3netCVPR}, which has achieved the state-of-the-art performance in music source separation.
Those architectures are illustrated in Fig.~\ref{fig:netarch}~(a)--(c).

The first variant is RD3Net~\cite{shimada2021accdoa}.
To adapt D3Net architecture to the SELD problem, three modifications are made. First, dense blocks in the up-sampling path is omitted because high frame-rate prediction is not necessary for the SELD problem. Second, the bottleneck part is replaced with gated recurrent unit~(GRU) cells. Third, the batch normalization is replaced with the network deconvolution~\cite{Ye2020}. 
The architecture is illustrated in Fig.~\ref{fig:netarch}~(a).

% \begin{comment}
% To increase the diversity of the model ensemble, we also consider two D3Net variants. 
% We propose replacing the GRU block with a time-frequency RNN~(TFRNN) block, which is inspired by a dual-path recurrent neural network~(DPRNN) block~\cite{Luo2020}. DPRNN was originally proposed to model long sequence. The long sequence is modeled by the intra- and inter-chunk RNN blocks (i.e., dual path).
% Each RNN block has the same architecture, which is composed of bidirectional long short-term memory~(LSTM), fully connected layers, and layer normalization. Both RNN blocks are alternately stacked.
% The dual path is considered to be able to  efficiently integrate two types of contexts. In TFRNN block, we utilize the dual-path architecture to integrate both time-domain context and frequency-domain context. 
% As shown in Fig.~\ref{fig:netarch}~(b), TFRNN with four RNN blocks is applied to the output of fourth dense block and followed by a frequency-domain pooling operation.
% % We refer to this architecture as time-frequency RNN (TFRNN). TFRNN is applied to the output of 4th dense block and followed by a frequency-domain pooling operation. There are four RNN blocks in this experiment.
% \end{comment}
As the second variant, we propose to replace the GRU block with a time-frequency RNN~(TFRNN) block, which is inspired by a dual-path recurrent neural network~(DPRNN) block~\cite{Luo2020}. 
DPRNN was originally proposed to model long sequence. 
The long sequence is modeled by the intra- and inter-chunk RNN blocks (i.e., dual path).
The dual path is considered to be able to efficiently integrate both time-domain context and frequency-domain context. 
Each RNN block is composed of bidirectional long short-term memory~(LSTM), fully connected layers, and layer normalization. 
The intra- and inter-chunk RNN blocks are alternately stacked.
% The dual path is considered to be able to  efficiently integrate two types of contexts. 
% In the TFRNN block, we utilize the dual-path architecture to integrate both time-domain context and frequency-domain context. 
As shown in Fig.~\ref{fig:netarch}~(b), TFRNN with four RNN blocks is applied to the output of fourth dense block, and followed by a frequency-domain pooling operation.

The last variant is inspired by the EINV2 architecture~\cite{cao2021improved}. 
Soft parameter-sharing using cross-stitch and a transformer encoder are incorporated into D3Net.
As shown in Fig.~\ref{fig:netarch}~(c), EINV2-D3Net consists of two parts, to which amplitude spectrograms features with and without IPDs are fed respectively.
% While a part get amplitude spectrograms and IPDs, the other part get only amplitude spectrograms.
Each part has four D3 blocks and one track-wise transformer encoder block. 
An integration layer combines the outputs of the two parts, and outputs ACCDOA vectors. 
Note that the differences from the original EINV2~\cite{cao2021improved} are twofold: 1) replacement of conv-block with D3 block and 2) the integration layer after the transformer encoder blocks.

\subsection{Post-processing}
\label{ssec:post}
% In the inference, the shift frame length is less than the input frame length, and the system outputs the average of each frame-wise output.
During the inference, we split the 60-second inputs into shorter segments that overlap.
Subsequently, each segment is processed, and the results of overlapped frames are averaged.
To further improve the performance, we conduct a post-processing by rotating the FOA data, estimating the ACCDOA vectors, rotating the vectors back, and averaging the vectors of different rotation patterns~\cite{shimada2021accdoa}.

\subsection{Hyper-parameters and training procedure}
\label{ssec:hyper}
The sampling frequency is set to 24 kHz. The STFT is used with a 20 ms frame length and 10 ms frame hop. The frame length of input to the networks is 512 frames. The frame shift length is set to 20 frames during the inference.
We use a batch size of~32. Each training sample is generated on-the-fly~\cite{erdogan2018investigations}. 
We gradually increase the learning rate to 0.001 with 50,000 iterations~\cite{goyal2017accurate}. After the warm-up, the learning rate is decreased by 10\% if the SELD score of the validation do not improve in 40,000 consecutive iterations. We use the Adam optimizer with a weight decay of~$10^{-6}$.
We validate and save model weights every 10,000 iterations up to 400,000 iterations. Finally we average the model weights from the last 10 models as in \cite{karita2019comparative}.

\section{EINV2-based System}
\label{sec:einv2_system}
The EINV2 framework uses the track-wise output format and permutation-invariant training~(PIT)~\cite{cao2021improved}.
% The track-wise output format can detect DOAs using only necessary dimensions. 
% It can also detect overlaps of the same class.
The track-wise format assumes that the output of the model has several tracks, each with at most one predicted event with a corresponding DOA.
Different tracks can detect events of the same class with different DOA, which enables us to detect overlaps of the same class.
% The major difference from ACCDOA framework is that EINV2 is optimized with multiple output targets using PIT by solving the track permutation problem, while ACCDOA is optimized with a single target criterion.
EINV2 is optimized with multiple targets using PIT by solving the track permutation problem.
We use a binary cross entropy~(BCE) for the SED classification task and a MSE for the DOA regression task~\cite{cao2021improved}. 

% \subsection{Other modules}
% \label{ssec:einv2_others}
% Here, the modules such as features, data augmentation, post-processing, hyper-parameters, and training procedure in the EINV2-based system are the same as in the ACCDOA-based systems.
The other components such as features, data augmentation, post-processing, hyper-parameters, and training procedure in the EINV2-based system are the same as in the ACCDOA-based systems.
As shown in Fig.~\ref{fig:netarch}~(d), the network architecture use soft parameter-sharing using cross-stitch and a transformer encoder, the same as the original EINV2 architecture~\cite{cao2021improved} except for replacement of conv-block with D3 block.

\section{Model ensemble}
\label{sec:ensemble}

\begin{table}[t]
    \centering
    \caption{Ensemble configuration. Th. means threshold.}
    \vspace{1mm}
        \begin{tabular}{l|c|c|l}
        \toprule
        Ens. & Avg. & Th. & Base system \\
        \midrule
        \multirow{4}{*}{\#1} & & \multirow{4}{*}{0.3} & ACCDOA-based RD3Net $\times$11 \\
        & Simple & & ACCDOA-based TFR-D3Net \\
        & avg.   & & ACCDOA-based EINV2-D3Net \\
        &        & & EINV2-based  EINV2-D3Net $\times$2 \\
        \midrule
        \multirow{4}{*}{\#2} & & \multirow{4}{*}{0.4} & ACCDOA-based RD3Net $\times$11 \\
        & Simple & & ACCDOA-based TFR-D3Net \\
        & avg.   & & ACCDOA-based EINV2-D3Net \\
        &        & & EINV2-based  EINV2-D3Net $\times$2 \\
        \midrule
        \multirow{4}{*}{\#3} & & \multirow{4}{*}{0.3} & ACCDOA-based RD3Net $\times$11 \\
        & Weighted & & ACCDOA-based TFR-D3Net \\
        & avg.     & & ACCDOA-based EINV2-D3Net \\
        &          & & EINV2-based  EINV2-D3Net $\times$2 \\
        \midrule
        \multirow{4}{*}{\#4} & & \multirow{4}{*}{0.4} & ACCDOA-based RD3Net $\times$15 \\
        & Weighted & & ACCDOA-based TFR-D3Net $\times$2 \\
        & avg.     & & ACCDOA-based EINV2-D3Net $\times$2 \\
        &          & & EINV2-based  EINV2-D3Net $\times$4 \\
        \bottomrule
        \end{tabular}
    \vspace{-3mm}
    \label{tb:ensemble}
\end{table}

\begin{table*}[t]
    \centering
    \caption{SELD performance of our systems evaluated by using joint localization/detection metrics for the development set.}
    \vspace{1mm}
        \begin{tabular}{l|c|cccc|cccc}
        \toprule
        & Number of & \multicolumn{4}{c|}{Validation fold} & \multicolumn{4}{c}{Testing fold} \\
        System & parameters & ${ER}_{20^{\circ}}$ & ${F}_{20^{\circ}}$ & ${LE}_{CD}$ & ${LR}_{CD}$ & ${ER}_{20^{\circ}}$ & ${F}_{20^{\circ}}$ & ${LE}_{CD}$ & ${LR}_{CD}$ \\
        \midrule
        Baseline FOA~\cite{politis2021dataset} & 0.5 M & - & - & - & - & 0.73 & 30.7 & $24.5^{\circ}$ & 40.5 \\
        % \midrule
        ACCDOA-based RD3Net             & 1.7 M & 0.44 & 68.1 & $14.3^{\circ}$ & 70.0 & 0.48 & 64.1 & $13.2^{\circ}$ & 63.2 \\  % E1.1 OS
        ACCDOA-based TFR-D3Net   w/ IRS & 3.0 M & 0.42 & 70.2 & $12.0^{\circ}$ & 70.5 & 0.45 & 67.5 & $11.5^{\circ}$ & 67.0 \\  % E2.1 OS
        ACCDOA-based EINV2-D3Net w/ IRS & 7.1 M & 0.43 & 69.2 & $12.0^{\circ}$ & 71.7 & 0.44 & 67.9 & $11.5^{\circ}$ & 69.5 \\  % E2.4 OS
        EINV2-based  EINV2-D3Net w/ IRS & 7.1 M & 0.35 & 75.9 & $12.3^{\circ}$ & 79.2 & 0.48 & 63.5 & $11.5^{\circ}$ & 64.9 \\  % E2.3 OS
        \midrule
        Ensemble \#1 & 42.6 M & 0.39 & 75.3 & $11.5^{\circ}$ & 82.6 & 0.43 & 69.6 & $11.3^{\circ}$ & 73.2 \\  % Ens1
        Ensemble \#2 & 42.6 M & 0.36 & 75.7 & $10.9^{\circ}$ & 77.6 & 0.42 & 69.6 & $10.7^{\circ}$ & 68.6 \\  % Ens2
        Ensemble \#3 & 42.6 M & 0.38 & 75.8 & $11.3^{\circ}$ & 83.0 & 0.43 & 69.9 & $11.1^{\circ}$ & 73.2 \\  % Ens3
        Ensemble \#4 & 73.6 M & 0.35 & 75.9 & $10.6^{\circ}$ & 77.4 & 0.41 & 70.0 & $10.3^{\circ}$ & 68.7 \\  % Ens4
        \bottomrule
        \end{tabular}
    \label{tb:result}
\end{table*}

\begin{table}[t]
    \centering
    \caption{Our preliminary experimental result without and with IRS for the development set.}
    \vspace{1mm}
        \begin{tabular}{l|cccc}
        \toprule
        & \multicolumn{4}{c}{Testing fold} \\
        ACCDOA-based system & ${ER}_{20^{\circ}}$ & ${F}_{20^{\circ}}$ & ${LE}_{CD}$ & ${LR}_{CD}$ \\
        \midrule
        Without IRS & 0.55 & 56.6 & $16.1^{\circ}$ & 57.6 \\  % Trial New Metrics 15.0
        With IRS    & 0.54 & 57.6 & $15.2^{\circ}$ & 58.5 \\  % Trial New Metrics 15.1
        \bottomrule
        \end{tabular}
    \label{tb:result_irs}
\end{table}

A model ensemble is performed by averaging outputs of several models trained with different conditions such as input features, training folds, and model architectures.
We convert the EINV2 format into the ACCDOA format for averaging by the following two steps: making pseudo track-wise ACCDOA outputs by multiplying SED outputs and DOA output for each track, and taking non track-wise ACCDOA outputs with the maximum activity among the tracks.

Here we use the simple average and the weighted average of the outputs predicted by different model.
The weights are assigned to each model, thus the dimension of weights is $M$, where $M$ is the number of models.
We use 0.3 or 0.4 as a threshold.
During the inference, if the norm of the ACCDOA vector of a sound event class exceeds the threshold, we consider the class to be active.
The models used for the ensemble are listed in Table~\ref{tb:ensemble}.

\section{Experimental evaluation}
\label{sec:eval}

In this section, we show the experimental settings and our results on the development dataset.

\subsection{Experimental settings}
\label{ssec:setting}
We evaluated our approach on the development set of TAU Spatial Sound Events 2021 - Ambisonic using the suggested setup~\cite{politis2021dataset}. 
The baseline was an ACCDOA-based system with a convolutional recurrent neural network (CRNN)~\cite{politis2021dataset}.
In the setup, four metrics were used for the evaluation~\cite{mesaros2019joint}. The first was the localization error ${LE}_{CD}$, which expresses the average angular distance between predictions and references of the same class. The second was a simple localization recall metric ${LR}_{CD}$, which expresses the true positive rate of how many of these localization predictions were correctly detected in a class out of the total number of class instances. The next two metrics were the location-dependent error rate~(${ER}_{20^{\circ}}$) and F-score~(${F}_{20^{\circ}}$), where predictions are considered as true positives only when the distance from the reference is less than  $20^{\circ}$.
% considering that true positives were predicted only under a distance threshold of $20^{\circ}$ from the reference.

\subsection{Experimental results}
\label{ssec:result}

Table~\ref{tb:result} shows the performance with the development set for our systems. As shown in the table, our ACCDOA-based RD3Net outperformed the baseline for each metric by a large margin.
The result also suggested that D3Net variants and IRS data augmentation improved performance.
Model ensemble improved ${F}_{20^{\circ}}$ by 2.1 points from the single model in the testing fold.
The result showed that model ensembles constantly performed better than single models.
There were no significant difference between the simple average and the weighted average.

Table~\ref{tb:result_irs} shows our preliminary experiment result of an ACCDOA-based system without and with IRS. We observed that the performance with IRS was better than without IRS.

\section{Conclusion}
\label{sec:concl}
We presented our approach to DCASE2021 task 3: sound event localization and detection~(SELD) with directional interference. Our systems use the activity-coupled Cartesian direction of arrival~(ACCDOA) representation to solve both sound event detection~(SED) and DOA estimation tasks in a unified manner. Using the ACCDOA-based system with RD3Net as a base, we perform model ensembles by averaging outputs of several systems trained with different conditions such as input features, training folds, and model architectures. We also use the event independent network v2~(EINV2)-based system to increase the diversity of the model ensembles. To generalize the models, we further propose impulse response simulation~(IRS), which generates simulated multi-channel signals by convolving simulated room impulse responses~(RIRs) with source signals extracted from the original dataset. Our systems performed superiorly to the baselines without model ensemble. Furthermore, we observed further improvement with ensembles of the ACCDOA- and EINV2-based systems.

% -------------------------------------------------------------------------
% Either list references using the bibliography style file IEEEtran.bst
\bibliographystyle{IEEEtran}
\bibliography{refs}

\end{sloppy}
\end{document}